\documentclass[pre,aps,amsmath,amssymb]{revtex4}

\usepackage{graphicx}
\usepackage{bm}

\newcommand{\beq}{\begin{equation}}
\newcommand{\eeq}{\end{equation}}
\newcommand{\bea}{\begin{eqnarray}}
\newcommand{\eea}{\end{eqnarray}}
\newcommand{\re}{\text{Re}}
\newcommand{\im}{\text{Im}}
\newcommand{\ie}{i.e. }
\newcommand{\eg}{e.g. }

\let\a=\alpha \let\b=\beta  \let\g=\gamma  \let\d=\delta 
\let\z=\zeta  \let\h=\eta     \let\l=\lambda
\let\m=\mu                 
\let\s=\sigma \let\t=\tau   \let\f=\varphi 
   
   \let\L=\Lambda 
    \let\Si=\Sigma     
\let\O=\Omega 

\let\io=\infty

\begin{document}


\title
{\large{\bf Glassy behavior of light in random lasers }}

\author{L. Angelani$^1$, C. Conti$^{2,3}$, G. Ruocco$^{3,4}$, and F. Zamponi$^{5}$}

\affiliation{
$^1$Research center SMC INFM-CNR, c/o Universit\`a
di Roma ``La Sapienza,'' I-00185, Roma, Italy \\
$^2$Centro studi e ricerche ``Enrico Fermi'', Via Panisperna 89/A,
I-00184, Roma, Italy   \\
$^3$Research center Soft INFM-CNR, c/o Universit\`a di
Roma ``La Sapienza,'' I-00185, Roma, Italy \\
$^4$Dipartimento di Fisica, Universit\`a di Roma ``La Sapienza,''
I-00185, Roma, Italy \\
$^5$Laboratoire de Physique Th\'eorique de l'\'Ecole Normale Sup\'erieure,
24 Rue Lhomond, 75231 Paris Cedex 05, France
}

\begin{abstract}
A theoretical analysis
[Angelani et al., Phys. Rev. Lett. {\bf 96}, 065702 (2006)]
predicts glassy behavior of light in 
a nonlinear random medium. 
This implies slow dynamics related to the presence of many
metastable states.
We consider very general equations (that also apply to other systems, like Bose-Condensed gases) 
describing light in a
disordered non-linear medium and through some approximations we relate them to
a mean-field spin-glass-like model. The model is solved by the replica method,
and replica-symmetry breaking phase transition is predicted. 
The transition describes a mode-locking process in which
the phases of the modes are locked to random (history and sample-dependent) values.
An extended discussion of possible experimental implications of our analysis is reported.
\end{abstract}


\maketitle


\section{Introduction}

In a nutshell, laser action in a stochastic resonator (SR) defines a random
laser (RL). Following the original Lethokov's article, \cite{Letokhov68}
a SR is a disordered medium sustaining a large number of electromagnetic modes with overlapping resonances.
The modes are not necessarily localized (in the Anderson sense), but can be extended modes in a random medium; they
have typically a finite life-time and are sometimes referred to as ``quasi-modes''.
Generally speaking, we will refer to a RL as a multi-mode laser system
that displays some disorder; this will be described by a probability
distribution and we will then consider different realizations of the system. Such
a general definition not only embraces the  experiments addressed below, but also include standard 
lasers with a disordered cavity, or integrated devices, as for example ordered photonic crystals \cite{SakodaBook} infiltrated by 
some active (i.e. doped) soft-material, like liquid crystals or polymers, that induces a given amount of disorder, or 
even intentionally disordered photonic crystals enriched by quantum wells providing optical gain.

In the early developments, the theoretical framework at the basis of RL 
has relied on light diffusion  \cite{Sha94,Lawandy94,Wiersma95}.
These studies stimulated many investigations concerning photon dynamics in a disordered medium,
up to considering the quantum transport of photons 
\cite{Beenakker98,Patra02, Hackenbroich03,Florescu04,Lodahl05,Storzer06,Skipetrov04}.
Subsequent detailed numerical 
studies revealed how important for RLs is the nature and the distribution of localized modes in random 
amplifying media, in particular in the strongly scattering
regime \cite{Jiang00,Vanneste02}.
Experiments were reported 
on the emerging of many coupled oscillation modes while increasing the pump energy and the consequent non-trivial
dynamics of the resulting optical signals \cite{Cao98,Cao99,Cao99b,Cao03}.
Coupling of modes were addressed in \cite{Cao03}, as the fact
that the maximum observed number of modes increases with the pumping
intensity and  with the
sample volume \cite{Ling01,Deych05}. 
Recent results pointed out new key issues concerning the physics of random lasers, as the
role of extended modes \cite{Mujumdar04,Song05}, or the presence of specific fluctuations.\cite{Anglos04,derMolen06}.

When considered from a semi-classical perspective, a multi-mode random laser strikingly displays those ingredients which are
typical of the physics of complexity: i.e. randomness and nonlinearity. The latter is due to typical mode-interaction processes,
like mode-competition and mode-locking \cite{HakenBook,Lamb64,Ducuing64,Bryan73,MeystreBook}.
Complex processes in laser physics, including nonlinear optics,  are well known and studied 
(see e.g. \cite{Arecchi99,BadiiBook}), 
up to recent investigations in multi-mode systems \cite{Papoff04,Cabrera2006, Sierks98} 
and successful reformulations of standard laser thermodynamics 
\cite{Gordon02,Gordon03,Gordon03b,Weill05,Vodonos04}. 
The extension of these approaches to RL immediately leads to the application of the statistical theory of disordered systems, 
of which spin glass theory is a paradigm \cite{MPVBook}, and which is the subject of the present manuscript.

We show that, in the presence of a large number of coupled modes (extended or not), a mode-locking (ML) process can be observed.
ML is related to the relative phases between resonant states, which in some cases become locked at the same value.
For a standard laser it can be realized 
by an active device, like an acusto-optic modulator, or can be self-starting as in the presence of a nonlinearly mediated
mode interaction \cite{Haus00}. 
We can expect that RL-ML appears 
when many modes are put into oscillations, 
and their amplitudes are clamped at the oscillation values, which are random variables.
The temporal dynamics of the emitted signal is indeed 
strongly related to the phases \cite{YarivBook}, given the fact that the mode
amplitudes vary on a much longer time scale.
The latter circumstance favors the consideration of the mode amplitudes as ``quenched'' (i.e. random but slowly varying) variables, and the 
phases are to be taken as the relevant dynamical variables.
The mode-locking process  in standard lasers is now recognized as a thermodynamic phase transition 
\cite{Gordon02,Gordon03,Gordon03b,Weill05,Vodonos04};
it is expected, therefore, that the mode-locking transition for a RL takes the form of a phase transition in a disordered system.

This manuscript follows a recent letter \cite{Angelani06}, and furnishes: extensive and new details on the
derivation of the analytical results (including the stability analysis that was previously not reported);  
a discussion of the underlying working hypotheses;  a discussion on the nature of the considered electromagnetic modes;
the analysis of possible experimental frameworks where glassy behavior of light can be observed.
The paper is structured as follows. In section II, we will review coupled mode theory in a dielectric resonator in the
presence of a nonlinear susceptibility; in section III we will specialize the approach by deriving the leading model for our analysis;
in section IV we will apply the methods used in spin glass theory to solve the
model; section V is focused on a discussion of the physical meaning of
our results, using real-world units, and of possible experimental setups; conclusions are drawn in section VI.
\section{Coupled mode theory equations}
The physical system under consideration is an open electromagnetic cavity supporting modes 
 at optical frequencies. 
The cavity is characterized by the presence of disorder; 
for example randomly structured dielectrics
in between a couple of mirrors, or a mirrorless system (e.g. a distribution
of dielectric particles) such that there is a sufficiently high refractive index contrast, so that localized modes
(which means modes belonging to the discrete spectrum of the eigenvalue problem given by the Maxwell equations, 
as detailed below) do exist.
This is the case, for example, of a disordered distribution of TiO$_2$ particles, of semiconductor powders in a liquid
or a glassy matrix, or of
a nanostructured microcavity filled by a randomly fluctuating 
material like liquid crystals or soft matter.
The localized modes supported by these systems can be very different, depending on the degree of localizations, e.g. they can be 
distributed over all the the dielectric sample 
(as those investigated for example in the experiments reported in \cite{Mujumdar04}) or correspond
to well localized states
(as those numerically analysed in \cite{Vanneste02}); this distinction can
influence the properties of the interaction between modes, leading
to interesting effects, as will
be discussed in the following. However the physical picture we will obtain
is expected to be independent of the details of the interaction.

Models for multimode nonlinear optical cavities have been largely reported in literature 
(see e.g. \cite{SakodaBook,HausBook,Glauber91}).
Typically they result into coupled equations for complex amplitudes,
which can be obtained using various and equivalent approaches. In order to fix the notation and for the benefit of  the
non-expert reader, here we will briefly report a derivation based on a multiple
scale approach.
The electromagnetic cavity [a dielectric resonator (DR)],
is described by a (static) refractive index profile $n({\bf r})$.
Such a kind of system may support the existence of resonance modes, which can be either localized or distributed in the system.
Maxwell's equations are written as
\begin{equation}
\begin{array}{l}
\nabla\times\mathbf{H}=\varepsilon_0 n^2(\mathbf{r}) \partial_t\mathbf{E}\\
\nabla\times\mathbf{E}=-\mu_0 \displaystyle \partial_t \mathbf{H}
\end{array}
\end{equation}
The electric and magnetic fields can be expanded in normal modes with
angular frequencies $\omega_n$ and eigenvalues ${\bf E}_n (\bf{r})$ and
${\bf H}_n (\bf{r})$ as:
\begin{equation}
\begin{array}{l}
\mathbf{E}=\re [\sum_n \mathbf{E}_n (\mathbf{r}) \exp(-i\omega_n t)]\\
\mathbf{H}=\re [\sum_n \mathbf{H}_n (\mathbf{r}) \exp(-i\omega_n t)]  
\end{array}
\end{equation}
The latter quantities satisfy the generalized eigenvalue problem
\begin{equation}
\mathcal{L}\mathcal{F}_s=\omega_s \mathcal{M}\mathcal{F}_s
\end{equation}
while being
\begin{equation}
\mathcal{L}=\left( 
\begin{array}{cc}
0 & i\nabla\times\\
-i\nabla\times & 0\\
\end{array}
\right)
\end{equation}
\begin{equation}
\mathcal{M}=\left( 
\begin{array}{cc}
\epsilon_0 n^2(\mathbf{r}) & 0 \\
0 & \mu_0\\
\end{array}
\right)
\end{equation}
and 
\begin{equation}
\mathcal{F}_s=\left( 
\begin{array}{c}
{\mathbf{E}}_s\\
{\mathbf{H}}_s
\end{array}
\right)\text{.}
\end{equation}

Given a volume $V$ much wider than the DR, over which periodical (Born-Von Karman) boundary conditions are posed, 
and introducing the complex valued scalar product
\begin{equation}
\left( \mathbf{A},\mathbf{B}\right)=\int_V \mathbf{A}^* \cdot \mathbf{B} dV
\end{equation}
it turns out that $\mathcal{L}$ and $\mathcal{M}$ are self-adjoint operators. As a result $\omega_n$ are 
real valued and the eigenvectors are orthogonal with weight $\mathcal{M}$. 
Furthermore, since $({\mathbf{E}}_n,{\mathbf{H}}_n )$ and 
$({\mathbf{E}}^*_n,-{\mathbf{H}}^*_n)$ correspond to the same
eigenvalue $\omega_n$, $\mathbf{E}_n$ can be taken as real valued.

The average electromagnetic energy for each un-normalized mode is given by
\begin{equation}
\mathcal{E}_s=\frac{1}{4}\int_V  \varepsilon_0 n^2(\mathbf{r}) |{\mathbf{E}}_s|^2+\mu_0 
|{\mathbf{H}}_s|^2 dV=
\frac{1}{4}\left(\mathcal{F}_s,\mathcal{M}\mathcal{F}_s\right)\ .
\end{equation} 
In the following the modes are normalized in such a way
\begin{equation}
\frac{1}{4}\left(\mathcal{F}_s,\mathcal{M}\mathcal{F}_q\right)=\delta_{sq}
\end{equation}
Next we consider the perturbed Maxwell equations in the presence of a nonlinear polarization $\mathbf{P}_{NL}$, such that the overall dielectric displacement
vector is given by $\mathbf{D}=\varepsilon_0 n^2(\mathbf{r})\mathbf{E}+\mathbf{P}_{NL}$ and
$\mathbf{J}=\partial_t \mathbf{P}_{NL}$ a generalized current:
\begin{equation}
\begin{array}{l}
\nabla\times\mathbf{\mathbf{H}}=\varepsilon_0 n^2(r) \partial_t\mathbf{E}+\eta \mathbf{J}\\
\nabla\times\mathbf{E}=-\mu_0 \displaystyle \partial_t \mathbf{H}\text{,}
\end{array}
\end{equation}
$\eta$ is a bookkeeping perturbation parameter to be set equal to one at the end of the derivation.
Our aim is to write the solution of the nonlinear Maxwell equations as a superposition of modes such that the leading order has the form
\begin{equation}
\begin{array}{l}
\mathbf{E}=\re [\sum_n \sqrt{\omega_n} a_n(t) {\mathbf{E}}_n (\mathbf{r}) \exp(-i\omega_n t)]\\
\mathbf{H}=\re [\sum_n \sqrt{\omega_n} a_n(t) {\mathbf{H}}_n (\mathbf{r})\exp(-i\omega_n t)]  
\end{array}
\end{equation}
and the complex amplitudes $a_s$ are such that the total energy stored in the DR is 
\begin{equation}
\mathcal{E} = \Sigma_m \mathcal{E}_m  = \Sigma_m \omega_m \vert a_m \vert^2\text{.}
\end{equation}

There are various techniques to derive the leading equations for the $a_s$, here we adopt the multiple scale method (see e.g. \cite{NayfehBook}).
The perturbative expansion is written as  (with obvious notation)
\begin{equation}
\begin{array}{l}
\mathbf{E}=\re \{ \sum_n [  \sqrt{\omega_n} a_n(\eta t,\eta^2 t,...) {\mathbf{E}}_n+\eta {\mathbf{E}}_n^{(1)}+...]\exp(-i\omega_n t) \}\\
\mathbf{H}=\re \{ \sum_n [  \sqrt{\omega_n} a_n(\eta t,\eta^2 t,...) {\mathbf{H}}_n+\eta {\mathbf{H}}_n^{(1)}+... ]\exp(-i\omega_n t) \}
\end{array}
\end{equation}
where the amplitudes are taken to be dependent on the slow scales $t_n=\eta^n t$, as the first and higher order corrections like ${\mathbf{E}}_n^{(1)}$,
the fastest scale is $t_0=t$ and the temporal derivatives are written as $\partial_t=\partial_{t_0}+\eta \partial_{t_1}+...$.Letting
\begin{equation}
\mathbf{P}_{NL}=\re[\sum_n {\mathbf{P}}_n(t_1,t_2,...)\exp(-i\omega_n t_0) ]
\end{equation}
and
\begin{equation}
\mathbf{J}=\re[\sum_n {\mathbf{J}}_n(t_1,t_2,...)\exp(-i\omega_n t_0)]=\partial_t \mathbf{P}_{NL}
\end{equation}
with
\begin{equation}
\begin{array}{l}
{\mathbf{P}}_n={\mathbf{P}}_n^{(0)}+\eta {\mathbf{P}}_n^{(1)}+...\\
{\mathbf{J}}_n={\mathbf{J}}_n^{(0)}+\eta {\mathbf{J}}_n^{(1)}+...
\end{array}
\end{equation}
it is
\begin{equation}
{\mathbf{J}}_n^{(0)}=-i\omega_n {\mathbf{P}}_n^{(0)}\text{.}
\end{equation}

Using the previous machinery into the nonlinear Maxwell equations at the first order in $\eta$ it is found for the 
term oscillating with $\exp(-i\omega_s t_0)$
\begin{equation}
\label{firstorder}
\mathcal{L}\mathcal{F}_s^{(1)}-\omega_s \mathcal{M}\mathcal{F}_s^{(1)}=\mathcal{B}_s
\end{equation}
while having
\begin{equation}
\mathcal{F}_s^{(1)}=\left( \begin{array}{c} {\mathbf{E}}_s^{(1)}\\ {\mathbf{H}}_s^{(1)} \end{array}\right)
\end{equation}
and
\begin{equation}
\mathcal{B}_s=\left( \begin{array}{c} 
i\varepsilon_0 n^2(\mathbf{r})\sqrt{\omega_s}\displaystyle\frac{d  a_s}{d t_1}{\mathbf{E}}_s+i {\mathbf{J}}_s^{(0)} \\ 
i\mu_0 \displaystyle\sqrt{\omega_s} \frac{d a_s}{d t_1}{\mathbf{H}}_s 
\end{array}\right)
\end{equation}

The Fredholm theorem applied to (\ref{firstorder}), states that the solvability condition is the orthogonality with
the kernel solution, i.e. $\mathcal{F}_s$: $(\mathcal{F}_s,\mathcal{B}_s)=0$. Hence
\begin{equation}
\sqrt{\omega_s}\frac{d  a_s}{d t_1}=\frac{i\omega_s}{4}\left({\mathbf{E}}_s, {\mathbf{P}}_s^{(0)}\right)\text{.}
\end{equation}
Going back to the original variables, we have the desired result
\begin{equation}
\label{coupledmodes2a} \frac{d
a_s(t)}{dt}=-\frac{\sqrt{\omega_s}}{4i} \int_V \mathbf{
E}_s^*(\mathbf{r}) \cdot \mathbf{ P}_s(\mathbf{r}) \;  dV \ .
\end{equation}

\section{Nonlinear susceptibility and mode interactions in active random cavities}

We consider the case in which many modes are put into oscillations and 
interact due to the nonlinearity of the amplifying medium. 
In resonant systems the nonlinear optical response
can be found from the density matrix equations in a two-level system, as originally investigated by
Lamb \cite{Lamb64}.
The component of the nonlinear susceptibility oscillating at $\omega_s$ is modelled as usual \cite{BoydBook} 
and is written as:
\begin{equation}
\label{NLP}
P_{s}^\alpha=\!\!\!\!\sum_{\omega_s+\omega_p=\omega_q+\omega_r}\!\!\!\! 
 \chi_{\alpha\beta\gamma\delta}
(\omega_s;\omega_q,\omega_r,-\omega_p,\mathbf{r})
 E_p^\beta( \mathbf{r}) E_q^\gamma ( \mathbf{r}) E_r^\delta( \mathbf{r}) \sqrt{ \omega_p \omega_q \omega_r} 
a_q a_r a_p^*
\end{equation}
where $\chi$ is the 
third order response susceptibility tensor, which in general depends on the positions in the DR. 
Using (\ref{NLP}) the coupled mode theory equations (\ref{coupledmodes2a}) read as
\begin{equation}
\label{NLP1}
\frac{d a_s}{dt}=-\frac{1}{2} \sum_{pqr} g_{spqr} a_q a_r a_p^* \ ,
\end{equation}
while being 
\begin{equation} \label{gs}
 g_{spqr}=\frac{\sqrt{\omega_s \omega_p \omega_q \omega_r}}{2i}\int_V 
\chi_{\alpha\beta\gamma\delta}
(\omega_s;\omega_q,\omega_r,-\omega_p,\mathbf{r}) E_s^\alpha( \mathbf{r})
E_p^\beta( \mathbf{r})  
E_q^\gamma( \mathbf{r})  E_r^\delta( \mathbf{r}) dV \ .
\end{equation}

\subsection{Mode interactions and the role of localized modes}
\label{sec:modes}
Our treatment follows early works on multimode cavities \cite{Lamb64,Bryan73}
and consistently, in the previous equations, intermode frequencies and higher
harmonics are neglected because 
they have in general a lower Q-factor if compared 
to those of the supported cavity modes.
Additionally, since the sum in Eq.~(\ref{NLP1}) is extended to all the modes combination satisfying the condition
$\omega_s=\omega_q+\omega_r-\omega_p$, we recall that the frequencies satisfying this relation can be divided into three categories \cite{Bryan73}:
(a) $\omega_s=\omega_q$ and $\omega_r=\omega_p$; (b) $\omega_s=\omega_r$ and $\omega_q=\omega_p$; and (c) 
$\omega_s=\omega_q+\omega_r-\omega_p$ excluding (a) and (b).
Categories (a) and (b) were shown to determine the oscillation values of the energies of the modes $\mathcal{E}_s$, and provide terms 
like self and cross-saturation, as also recently considered in \cite{Deych05}, with reference to RLs. 
The third group are the ``combination tone terms'' \cite{Bryan73} which were originally neglected, even if it was later recognized,
through numerical calculations,
to have a role when the number of modes increases \cite{Brunner83}.
We are interested to the regime in which a large number of modes is put into oscillation 
in a limited spectral range around a given carrier wavelength $\omega_0$ (which can be taken as the resonant angular frequency of the active medium)
 and we will show below that in RLs, the combination
tone terms provide a complex structure to the laser dynamics, as due to the fact that for an increasing number of modes they 
couple almost all the cavity resonances. 

In general, the resonant condition for the mode-locking processes $\omega_s=\omega_q+\omega_r-\omega_p$, does not need to be satisfied exactly but 
in such a way that the mode combination tone $\omega_q+\omega_r-\omega_p$ lies within the linewidth at $\omega_s$ 
(this is discussed e.g. in \cite{MeystreBook} with reference to three modes mode-locking). 
In the presence of many modes oscillating in a small bandwidth, and such that the linewidths are overlapping, as it is typically the
case for RLs (see the cited references) and (by definition) for SRs, many mode combination tones will couple to $\omega_s$, for which 
we have taken $\omega_s\cong\omega_q+\omega_r-\omega_p$ in (\ref{NLP1}). This opens the way
to a ``mean field theory'' where all the modes are coupled, \ie
the sum in (\ref{NLP1}) is over all the possible values of $pqr$.
We will describe this regime, moreover considering the thermodynamic limit as the number
of modes goes to infinity.

However, it is worth to observe that the coupling $g_{spqr}$ in (\ref{gs}) is
related to the spatial overlap of the four modes $E_s,E_p,E_q,E_r$ that enter
in the integral. In the case of {\it extended} modes, all the modes will have
large spatial overlap and $g_{spqr}\neq 0$ for all $spqr$, so the ``mean
field limit'' above is expected to be a very good approximation.
On the contrary, in the case of {\it strongly localized modes} with
localization length $\xi$, it is reasonable to
expect that only a finite number of modes will be supported in a localization volume
$\sim \xi^3$ \cite{Cao03,Vanneste02,Storzer06}, so that the coupling $g_{spqr}$ will be nonzero only for those modes
which are large in the same (or in adjacent) localization volumes. In this
case the interaction will be {\it short range}, \ie in the sum (\ref{NLP1})
only quadruplets of ``nearby'' modes will appear. Many intermediate situations
between the extended and the strongly localized 
ones might happen in random lasers
\cite{Cao03,Vanneste02,Storzer06,Mujumdar04} and indeed the precise nature of
the modes in these systems is not completely clear. 

In the short range case, the
basic phenomenology of the glass transition we will find (slow dynamics, random
mode-locking) remains the same, but the physics of the system is strongly
affected by {\it activated processes} (nucleation, barrier crossing, etc.)
which are negligible in the mean field limit. Indeed the nature of the glass
phase of short range spin glasses is still a debated problem \cite{YoungBook}.
Note that the localization length (and thus the interaction range) may vary on
many orders of magnitude and can be experimentally controlled
\cite{Storzer06}, at variance to what happens
in spin glasses and molecular or colloidal glasses, where the interaction
range is fixed by the property of the material and is always of the order of
the interparticle distance. 
This observation opens the
way toward the possibility of an experimental investigation of the crossover
between the mean field limit and the short range case that might be crucial
for the theoretical understanding of the glass phase in short range systems.

To summarize, we will assume that {\it i)} all the lasing modes have frequency
$\omega_s \cong \omega_0$, $\omega_0$ being the resonant frequency of the active medium,
so that the constraint $\omega_s = \omega_q + \omega_r - \omega_p$ can be released,
and that {\it ii)} the spatial overlap of the modes is large, so that the
integral in (\ref{gs}) will be not negligible for any quadruple of modes. Under these
hypotheses a ``mean field'' treatment in which all quadruples of modes
interact will be a very good approximation. Nevertheless, we expect the physical
picture we will find in the following to hold under much more general
assumptions on the interaction between modes. Its modifications due to the
violation of the hypotheses above will be very interesting for the theory of
spin glass systems.

\subsection{The ``quenched'' approximation: a Langevin equation for the phases}

Letting $a_s(t)=A_s(t) \exp[i\varphi_s(t)]$, 
 we take $A_s$ as slowly varying with
respect to $\varphi_s$.  Indeed, the facts that the temporal variation of the
 phases is on a time scale faster 
than that of the  amplitudes, and that fluctuations in a cavity take place because of the random
interference between modes and not because of the intensity
fluctuations of individual modes, 
are well established from the theory of mode-locking of standard 
multi-mode lasers  \cite{Ducuing64,HausBook,YarivBook}.
Previous analytical, or semi-analytical, studies \cite{Ducuing64,Lamb64,Bryan73}
(the RL case has been recently considered in \cite{Deych05})
 relayed on the so-called
``free run approximation'', i.e. the phases are taken to be rapidly varying and independent and can be averaged out (see Appendix \ref{App_0}).
This turns out into removing all phase-dependent terms in (\ref{NLP1}) and the resulting equations determine the 
amplitudes $A_s$, and hence the energy into each mode $\mathcal{E}_s$, which stays clamped at this value after
that the corresponding mode has been put into oscillation. 
As far as the phases can be taken as independent, the output laser signal displays small oscillations around an equilibrium value, 
because the noises into each mode amplitudes are independent. 
It is clear that in this approximation the combination tone terms in
(\ref{NLP1}) will disappear due to the averaging over the phases.
However, since the beginning \cite{Lamb64,Bryan73} (and later also confirmed
by detailed numerical investigations \cite{Brunner83})  
it has been known that this regime
holds as far as beating between modes, due to the mode combination tones, are negligible; and this is valid if a
few modes with nonoverlapping resonances are excited. Conversely, the mode combination terms are known to be responsible of ``mode locking'' 
processes, that in standard laser provide a fruitful approach to the generation of ultra-short pulses \cite{HausBook,YarivBook}.

Gain (described by an amplification coefficient $\gamma_s$) and radiation losses (measured by $\alpha_s$) 
are included in the equation of motion for the complex amplitudes
following a standard approach \cite{HausBook}:
\begin{equation}
\label{coupledmodes0}
\frac{d a_s}{dt}=-\frac{1}{2} \sum_{pqr} g_{spqr} a_q a_r a_p^*+(\g_s-
\a_s) a_s + \eta_s(t)  \ ,
\end{equation}
having introduced, as usual, a complex noise term, mainly due to
spontaneous emission (see e.g. \cite{GardinerBook,RiskenBook}),
with $\langle \h_p(t) \h_q(t')\rangle =\langle \h^*_p(t) \h^*_q(t')\rangle =0$
and $\langle \eta_p(t) \eta^*_q(t') \rangle=2 k_B T_{bath}
\delta_{pq}\delta(t-t')$, with $k_B$ the Boltzmann constant and
$T_{bath}$ an effective temperature, whose expression will be reported in a later section.
In (\ref{coupledmodes0}), the sum has been extended over all the modes, as
discussed above, and the contribution
of each possible combination tone to the amplitude $a_s$ is given by the relevant coupling coefficient $g_{spqr}$.

The tensor $g$ is a quantity symmetric with respect to the exchange of $s \leftrightarrow p$,
$q \leftrightarrow r$, while under $\{s,p\} \leftrightarrow \{q,r\}$ one has $g_{spqr}=g^*_{qrsp}$, 
see Eq.~(\ref{NLP}) and \cite{Bryan73,BoydBook}. 
Introducing the real-valued potential function
\begin{equation}
H=\frac{1}{4} \re \left[ \sum_{spqr} g_{spqr} a_q a_r a_p^* a_s^* \right]
=\frac14 \sum_{spqr} g^R_{spqr} a_q a_r a_p^* a_s^* - \frac1{4i} \sum_{spqr} g^I_{spqr} a_q a_r a_p^* a_s^* 
\ ,
\label{hfinal0}
\end{equation}
and letting $\mathcal{H}=\sum_s (\a_s-\g_s)|a_s|^2+H$,
the resulting model (\ref{coupledmodes0}) is re-written as
\begin{equation}
\frac{d a_s}{dt}=- \frac{\partial \mathcal{H}}{
\partial a^*_s} + \eta_s(t) \ ,
\end{equation}
where
\beq 
\frac{\partial}{\partial a^*} = \frac12 \left[ \frac{\partial}{\partial a^R} + i
  \frac{\partial}{\partial a^I} \right] \ .
\eeq
The previous equation can be cast in the form of a standard
Langevin equation for a system of $N$ ``particles'' moving in $2N$
dimensions (represented by $\{a^R_s,a^I_s \}_{s=1..N}$)
\cite{RiskenBook,Gordon03} and its invariant measure is given by
$\exp(-\mathcal{H}/k_B T_{bath})$.

The simplest case is attained when $g$ can be taken as real valued.
Indeed, considering Lamb theory for a two level system \cite{Lamb64}, which is the only approach providing an 
explicit expression for the susceptibility tensor $\chi$, one can show that the imaginary part
of $g$ vanishes as all the resonant frequencies are packed around a given value $\omega_0$.
The generalization to a complex $g$ is discussed in Appendix~\ref{App_0}.

Finally, the phases $\varphi_s$ can be taken as the relevant dynamic variables, 
due to the quenched approximation for the amplitudes $A_s$, see Appendix~\ref{App_0},
and the Hamiltonian  is written as
\begin{equation} \label{Ham}
{\cal H}(G,\varphi)={\cal H}_o + \sum_{\{sp\},\{qr\}} G_{spqr}
\cos(\varphi_s +\varphi_p -\varphi_q -\varphi_r)
\end{equation}
where ${\cal H}_o=\sum_s (\alpha_s-\gamma_s)A_s^2$ is an irrelevant constant term 
(as long as the amplitudes $A_s$ are constant)
and $G_{spqr}$=$2g_{spqr} A_s A_p A_q A_r$ is the real-valued coupling.
Note that the couplings $G_{spqr}$ are symmetric under internal permutations of the sets 
$\{s,p\}$ and $\{q,r\}$ and also under exchange $\{s,p\} \leftrightarrow
\{q,r\}$. Indeed the couplings have the same symmetry of the interaction term 
$\cos ( \f_{s}+\f_{p}-\f_{q}-\varphi_{r})$.
To count each term only once, the sum $\sum_{\{sp\},\{qr\}}$ in (\ref{Ham}) has been restricted
only to the values of $spqr$ which are
not related by the symmetries above, and correspondingly a factor of $8$ has been added in the 
coupling.

Hereafter we will consider these $G$ coefficients as
``quenched'' (due to the slow $t$ dependence of $A_s$), and the
relevant phase space is reduced to that spanned by $\varphi_s$.
The pump energy which controls the average energy into each mode (and hence the amplitudes $A_s$) fixes the 
amplitude of $G$.

\subsection{Gaussian random couplings}

If the cavity is realized by a random medium, as described above, the coupling coefficients $g$ 
are random variables, \ie they will depend on the specific sample one is considering. 
For a given cavity realization, the values of the coupling coefficients $g$
are determined by the specific nonlinear mechanism, \ie 
by the function $\chi$ in (\ref{NLP}), the mode frequencies $\omega_s$
and amplitudes $A_s$ and by the mode profiles $E_s(\mathbf{r})$ as expressed
in (\ref{gs}). All these are sources of randomness in the computation of $g$.
However, as we cannot compute the properties of the model for a specific
choice of the $g$, we will assume that the couplings are drawn from a given
probability distribution and compute average properties of the system with
respect to this distribution. It is possible to show that, in the
thermodynamic limit, these average quantities will be representative of many
of the properties of a given sample (\eg the free energy), see the discussion
in next sections and \cite{MPVBook}.

Given the fact that the fields are real-valued functions with positive and negative values 
at each point in the medium, and
additionally, parity of the modes may eventually make some coupling vanishing,
a possible choice, in order to simplify the problem as much as possible, 
is to consider the signs of these coupling coefficient (those corresponding to the
mode-combination tones) as random and 
treat them as Gaussian independent variables with zero mean, as detailed below.
This choice is further supported by the fact that, in general, the mode frequencies are symmetrically distributed with respect
to the resonant frequency $\omega_0$, and correspondingly the sign of the
nonlinear susceptibility largely varies \cite{BoydBook}.
The hypothesis of zero mean can be removed by generalizing the treatment
reported below following \cite{MPVBook}, leading to a very rich
phase diagram \cite{Gillin01}; different distributions of
the couplings can also be investigated but the problem becomes more difficult.

Additionally, it is important to point out the scaling properties of the Hamiltonian (\ref{Ham}).
Recalling that $E^\a_s = O(V^{-1/2})$ (due to
the normalization) and 
 $\chi_{\alpha\beta\gamma\delta}=O(1)$ are random variables, one has, 
from Eq.~(\ref{gs}), $g \sim V^{-2} \int_V R(\mathbf{r})
 dV \sim V^{-3/2}$, as $R(\mathbf{r})=V^2 \chi E \, E \, E \, E$ is an
 $O(1)$ random variable whose integral scales as $V^{1/2}$.
The coupling $G$s then scale as $\langle A^2 \rangle^2 g_0 V^{-3/2}$.
By a simple rescaling, the invariant measure can be
written as $\exp[-\beta H(J,\varphi)]$, where
$J_{pqrs}=G_{spqr}/ (g_0 \langle A^2 \rangle^2)$ has standard deviation
$1/V^{3/2} \propto 1/N^{3/2}$, as the number of modes is proportional to the
volume of the cavity, see \eg \cite{Cao03}: then, conventionally we will set
$\langle J^2 \rangle = 8/N^3$ and include all the system-dependent constants in
the definition of $\b$. 
Note that this scaling of the $J$s
guarantees that the Hamiltonian is extensive, \ie the average energy is
proportional to volume \cite{MPVBook,CC05}.
The parameter that controls the phase
transition is then $\beta=1/T = {\cal P}^2 /k_B T_{bath} $,
where ${\cal P}^2 = \langle
A^2 \rangle^2 g_0$. 
We recall that  $\omega_0 \langle A^2 \rangle$ measures the 
average energy per mode, while $g_0$ is a material-dependent constant.
Then ${\cal P}$ is proportional to the energy stored 
on average into each mode.
Hence, the relevant parameter for the lasers model is the 
adimensional ``{\it temperature}'' $T$: lowering $T$ can be obtained 
both lowering the bath temperature $T_{bath}$ (e.g. acting on the noise, as done
in recent experiments on the thermodynamics of standard lasers \cite{Gordon03b}) or increasing ``the 
pumping rate'' ${\cal P}$.
\section{Replica analysis of the model}
\label{replica_section}

Our interest here is to draw a mean field statistical description for random
lasers, which enables to go 
beyond the ``free run approximation'' and 
unveil the complex structures of the states of these systems, 
due to ``random mode-locking'' processes. We can do this by computing the
thermodynamic properties of the Hamiltonian (\ref{Ham}) describing the
stationary states of the system.

Summing up,
the random laser studied in the above sections is described by the disordered
mean-field Hamiltonian
\bea
 H = \sum_{\{pr\},\{sp\}} J_{spqr}
 \cos ( \f_{s}+\f_{p}-\f_{q}-\varphi_{r}) 
\label{H4}
\eea
where $\{\varphi_s\}$ are angular variables, 
$\varphi_s \in [0,2\pi)$, and 
$J_{spqr}$ are independent Gaussian random variables
with zero mean and variance $\overline{J^2} =\s_J^2 = 8/N^3$.
The sum in (\ref{H4}) is restricted to the values of $spqr$ that are not
related by the symmetries of the interaction term, see the discussion after
(\ref{Ham}).
Our purpose is to study the thermodynamics of this model.
In particular we are interested in average properties of the model
considering the variables $J$s as {\it quenched}: that
means, we will average the free energy, and not the partition function, over
the distribution of the couplings: averaging the partition function correspond
to considering the $J$s as dynamical variables evolving on the same time scale
of the phases. The average of the free energy can be
done by means of the replica trick \cite{MPVBook}. 
Here we report the calculation
in full detail, even if it is very similar to the replica calculation for the $p$-spin model,
see \eg \cite{CC05}.

\subsection{Replicated partition function}

The partition function is 
\begin{equation}
Z_N(\b,J) = \int d\{\varphi\} e^{-\beta H(J,\varphi)}\text{,}
\end{equation}
with free energy
\begin{equation} 
f_N(\b,J)= -\frac{T}N \ln Z_N(\b,J) \ .
\end{equation}
We want to calculate the free energy averaged over the disorder
\begin{equation}\label{freeenergy}
-\beta f(\b)=\lim_{N \rightarrow \infty} \frac{1}{N} \overline{\ln Z_N(\b,J)}=
 \lim_{N \rightarrow \infty} \lim_{n \rightarrow 0} \frac{\overline{[Z_N(\b,J)]^n}-1}{nN} \ ,
\end{equation}
where the overbar denotes the average over the random coupling $J$s
and, as usual in the replica method, one uses the formula 
$\ln Z = \lim_{n \to 0} (Z^n-1)/n$, introducing
the partition function of $n$ 
independent replicas $Z^n(J)\equiv [Z_N(\b,J)]^n$ with the same random couplings $J$s.
It is possible to show \cite{MPVBook} that the free energy is {\it
  self-averaging}, \ie one has
\beq
\lim_{N\to\io} f_N(\b,J) = f(\b)
\eeq
with probability one with respect to the distribution of the couplings $J$s; in
other words, in the thermodynamic limit the free energy of a given sample is
given, with probability $1$, by the average of the free energy over the
disorder, that we are able to compute.

In the following we will neglect all the multiplicative constants growing as
powers of $N$ in the partition function as they do not contribute to the free energy.
The $J$s have distribution $P(J)=\sqrt{N^3/16\pi} \exp{(-J^2N^3 / 16)}$:
by the relation
\begin{equation}
\int dJ P(J) e^{AJ} = \exp \left[\frac{4}{N^3}  A^2\right] \ ,
\end{equation}
for integer $n$, one has
\beq
\overline{Z^n(J)}=\int \left(\prod_{a=1}^{n} d\{\varphi^a \} \right) 
e^{\frac{\beta^2}{2} H_{eff}(\varphi^a_i)} \ , 
\eeq
with
\beq\begin{split}
H_{eff}&=\frac{8}{N^3} \sum_{a,b} \sum_{\{sp\},\{qr\}} 
\cos(\varphi^a_{s}+\varphi^a_{p}-\varphi^a_{q}-\varphi^a_{r}) 
\cos(\varphi^b_{s}+\varphi^b_{p}-\varphi^b_{q}-\varphi^b_{r}) \\
&=\frac{8N^{-3}}{2 \left(\frac{4}{2}!\right)^2} \sum_{a,b} \sum_{spqr}^{1,N} 
\cos(\varphi^a_{s}+\varphi^a_{p}-\varphi^a_{q}-\varphi^a_{r}) 
\cos(\varphi^b_{s}+\varphi^b_{p}-\varphi^b_{q}-\varphi^b_{r}) \\
&=\frac{N}{2} \sum_{a,b} \big[ |Q_{ab}|^4+|R_{ab}|^4 \big] \ ,
\end{split}\eeq
where in the second line the constraints on the sets $\{s,p\},\{q,r\}$ have been released.
In the above expression we have introduced the quantities:
\begin{gather}
Q_{ab}=N^{-1} \sum_i e^{i(\varphi_i^a-\varphi_i^b)} \ , \\
R_{ab}=N^{-1} \sum_i e^{i(\varphi_i^a+\varphi_i^b)} \ .
\end{gather}
Note that $Q_{aa} \equiv 1$ and $Q_{ba} = Q^*_{ab}$, while $R_{ba}=R_{ab}$. 
Also note that the effective Hamiltonian $H_{eff}$ depends only on the global variables $Q_{ab}$ and $R_{ab}$.

Using the notation $\delta(z)=\delta(z^R)\delta(z^I)$, the partition function can be 
written as
\begin{equation}
\overline{Z^n(J)}=\int \prod_{a > b} dq_{ab} \prod_{a\geq b} dr_{ab} \
e^{\frac{\beta^2}{2} H_{eff}(q_{ab},r_{ab})}  
\int \prod_{a=1}^{n} d\{\varphi^a \}
\prod_{a > b} \delta(q_{ab}-Q_{ab}) \prod_{a\geq b} \delta(r_{ab}-R_{ab}) \ ,
\end{equation} 
and the second integral can be easily evaluated introducing the integral representation for 
the complex $\delta$-function
\begin{equation}
\delta(z)=\int \frac{d\l}{(2\pi)^2} e^{\re( z \l^*)} \ , \ d\l=d\l^R \, d\l^I 
\end{equation}
and the integral is done on the imaginary axis of the complex $\l^R$, $\l^I$
planes (\ie one has to consider both $\l^R$ and $\l^I$ as complex numbers).
With some algebra one gets, with the convention that $[ab] \rightarrow a \geq b$ 
and $(ab) \rightarrow a>b$, and summing over the repeated indexes,
\begin{equation}
\begin{split}
&\int d\{\varphi^a \} \ \prod_{(ab)} \delta(q_{(ab)}-Q_{(ab)}) \prod_{[ab]} \delta(r_{[ab]}-R_{[ab]}) = \\
&\int d\l_{(ab)} d\m_{[ab]} \exp \left[ N \re( \l^*_{(ab)} q_{(ab)} + \m^*_{[ab]} r_{[ab]}) + 
N\ln Z(\l_{(ab)},\m_{[ab]}) \right] \ ,
\end{split}
\end{equation}
where
\begin{equation}
Z(\l_{(ab)},\m_{[ab]} )=\int d[\varphi^a] \exp 
\left[-\re \left( \l^*_{(ab)} e^{i(\varphi^a-\varphi^b)} +
\m^*_{[ab]} e^{i(\varphi^a+\varphi^b)} \right) 
\right] \ .
\end{equation}
(note the difference between $d\{\f^a\} = \prod_{a,i} d\f_i^a$ and $d[\f^a] =
\prod_a d\f^a$).
The partition function has then the form
\begin{equation}\label{ZSZ}
\overline{Z^n(J)}=\int  dq_{(ab)} d\l_{(ab)}dr_{[ab]} d\m_{[ab]} \exp \left[- N h(q,\l,r,\m) \right] \ ,
\end{equation}
where the function $h$ is given by
\begin{equation}
\begin{split}
\label{g-generale}
h(q,\l,r,\m)=&-\frac{\beta^2}{4} \left[ \sum_{a} |r_{aa}|^4 + n + 
2 \sum_{(ab)}\Big[ |q_{ab}|^4 + |r_{ab}|^4\Big] \right]  \\
-&\re(\m^*_{aa} r_{aa} +\l^*_{(ab)} q_{(ab)} +\m^*_{(ab)} r_{(ab)}) - 
\ln Z(\l_{(ab)},\m_{[ab]}) \ .
\end{split}
\end{equation}
We have extracted the diagonal part in the effective Hamiltonian reminding that 
$q_{aa} \equiv 1$ is fixed (this is why we didn't include the relative $\delta$-function).

The integral (\ref{ZSZ}) can be evaluated at the saddle-point.
The derivatives with respect to $q$ and $r$ yield
\begin{equation}
\begin{split}
&\l_{(ab)}=-2 \beta^2  |q_{(ab)}|^{2} \, q_{(ab)} \ , \\
&\m_{aa}=-\beta^2 |r_{aa}|^{2} \, r_{aa}  \ , \\
&\m_{(ab)}=-2 \beta^2 |r_{(ab)}|^{2} \, r_{(ab)} \ .
\end{split}
\end{equation}
Substituting these equations into (\ref{g-generale}), $h$ is
\begin{equation}
\label{g-generale2}
h(q,r)=- \frac{n\b^2}{4} +  \frac{3\beta^2}{4}
 \left[ \sum_{a} |r_{aa}|^4  + 
2 \sum_{(ab)} |q_{ab}|^4 + |r_{ab}|^4 \right]
 - \ln Z(q_{(ab)},r_{[ab]}) \ .
\end{equation}
To perform the analytic continuation to $n\to 0$ one has to
make an {\it ansatz} on the structure of the matrices $q$ and $r$. 
The free energy is then computed using the equation (\ref{freeenergy}).
Using the relation
\begin{equation}
\lim_{N \rightarrow \infty} \overline{Z^n} \sim e^{-N \text{min}[h(q,r)]}
\end{equation}
and assuming that $\text{min}[h(q,r)] \sim n$ (\ie it is small) we have for the free energy
\begin{equation}
\beta f = -\lim_{n \rightarrow 0} \frac{e^{-N\min[h(q,r)]}-1}{nN} \sim 
\lim_{n \rightarrow 0} \frac{N\min[h(q,r)]}{nN} = 
\text{min} \left[ \lim_{n \rightarrow 0} n^{-1}h(q,r) \right] = \min [\b\phi(q,r)] \ .
\end{equation}
Note that the limits $n\to 0$ and $N\to \io$ have been exchanged \cite{MPVBook}.
A reasonable {\it ansatz} that we will make is that $r_{ab} \equiv 0$ at the
saddle point. Indeed, we are looking for disordered states that are usually
characterized by a non-zero overlap $q$ and a vanishing magnetization, \ie the
rotational symmetry is not broken.
As $r$ is not invariant under rotations,
we will set $r=0$ in the following. Then we have to minimize
\beq\label{f2min}
\begin{split}
&\b\phi(q)= -\frac{\b^2}{4} + \frac{3\beta^2}{2n}
\sum_{a > b} |q_{ab}|^4
 - n^{-1}\ln Z(q) \ , \\
&Z(q) = \int d\varphi^a \exp 
\left[\re \sum_{a>b} 2 \b^2 |q_{ab}|^{2} q^*_{ab} e^{i(\varphi^a-\varphi^b)}
\right] \ ,
\end{split}
\eeq
for $n\to 0$. This function is very similar to the one that describes the Ising $p$-spin glass.

The vanishing of the derivative with respect to $q_{ab}$ of $\b \phi(q)$ gives the saddle point equation
\beq
q_{ab} = \langle e^{i(\f^a-\f^b)} \rangle \ ,
\label{spe}
\eeq
where the average is on the measure that defines $Z(q)$.
Indeed, performing the derivative with respect to the real and imaginary part of $q_{ab}$, one obtains 
two equations that can be written in a single saddle point equation for $q_{ab}$
\beq\label{spe1}
3\ q_{ab} = \langle e^{i(\f^a-\f^b)} \rangle + 2 \frac{q_{ab}}{|q_{ab}|^2} 
\ \re\  q^*_{ab} \ \langle e^{i(\varphi^a-\varphi^b)}\rangle \ .
\eeq
It is easy to show that Eq.~(\ref{spe})
is solution of Eq.~(\ref{spe1}).

The replica symmetric solution corresponds to $q_{ab}\equiv q$; in particular $q=0$ is a
solution of the saddle point equations and is the stable one in the high
temperature phase. Another solution appear at low temperatures but is always
unstable (see below).
The {\sc rs} free energy is simply $f_{RS} = -\b/4$ as in the Ising $p$-spin
glass (neglecting irrelevant constants).
\subsection{One step replica symmetry breaking} 
The {\sc 1rsb} {\it ansatz} is the following:
we divide the matrix $q_{ab}$ in $n/m$ blocks of side $m$. 
The elements in the off-diagonal blocks are set to $0$ while in the diagonal blocks {\sc rs}
is assumed and $q_{ab}=q$. The simplest choice is to assume that $q$ is
real. This is very reasonable due
to the rotational symmetry, see \eg the discussion in \cite{BiYo86}, p.~894,
and moreover in this way the constraint $q_{ab}=q_{ba}^*$ is respected.
For instance, we have for $n=6$ and
$m=3$:
\begin{equation} \label{QQ1rsb}
\big(q_{ab}\big)=\left(
\begin{array}{cc}
\left(
\begin{array}{ccc}
1 & q & q \\
q & 1 & q \\
q & q & 1 \\
\end{array}
\right) & 0 \\
0 & 
\left(
\begin{array}{ccc}
1 & q & q \\
q & 1 & q \\
q & q & 1 \\
\end{array}
\right) \\
\end{array}
\right)
\end{equation}
Then one has
\beq
\lim_{n\to 0} n^{-1} \sum_{a>b} q_{ab}^4 =\frac{1}{2} (m-1) q^4 \ ,
\eeq
and, as the variables $\f^a$ in different blocks become uncorrelated,
\beq\begin{split}
Z(q)&=\left[  \int \prod_{a=1}^m d\varphi^a
e^{\b^2 q^{3} \sum_{a \neq b} e^{i(\varphi^a-\varphi^b)}} \right]^{n/m} 
=\left[  \int \prod_{a=1}^m d\varphi^a
e^{\b^2 q^{3} \left[ \left|\sum_{a} e^{i\varphi^a}\right|^2 - m 
\right]  } \right]^{n/m} \\
&=\left[  e^{-m\b^2 q^{3}}
\int \prod_{a=1}^m d\varphi^a
\int {\cal D} \z e^{\b \sqrt{2 q^3} \re \z^* \sum_a e^{i\f^a}}
 \right]^{n/m} \ ,
\end{split}
\eeq
where $\z$ is a complex variable and ${\cal D}\z = \frac{d\re\z \, 
d\im\z}{2\pi}e^{-\frac{1}{2} \z \z^*}$.
Defining also $\l = \sqrt{2 q^3}$ one has
\beq
n^{-1} \ln Z(q) =  -\b^2 q^3 +
m^{-1} \ln \int {\cal D}\z \left( \int d\f e^{\b \l \re \z^* e^{i\f}} \right)^m \ .
\eeq
Introducing the modified Bessel function of the first kind of order $0$
\beq
I_0(\b \l |\z|) =\frac{1}{2\pi} \int_0^{2\pi} d\f \, e^{\b \l \re \z^* e^{i\f}} \ ,
\eeq
and noting that it depends on the modulus $z=|\zeta|$
one finally gets (apart from constant terms)
\beq
\b\phi_{1RSB}(m,T) = 
 -\frac{\b^2}{4}\big[ 1 + 3 (1-m) q^4
 - 4 q^3 \big] - \frac{1}{m} \ln \int_0^{\infty} {\cal D}z \ I_0^m(\b\l z) \ ,
\label{f1rsb}
\eeq
where ${\cal D}z= z e^{-z^2/2} dz$. The value of $q$ is determined by the 
condition $\partial_q \phi_{1RSB}=0$ that gives (see Appendix \ref{App_B} for details):
\beq
q=\frac{\int_0^{\infty} {\cal D}z I_0^m(\b\l z) \left[\frac{I_1(\b\l z)}{I_0(\b\l z)}\right]^2}
{\int_0^{\infty} {\cal D}z I_0^m(\b \l z)} \ ,
\label{q1rsb}
\eeq
where $I_1(x)=I_0'(x)$ is the modified Bessel function of order $1$.
This expression is similar to the {\sc 1rsb} free energy for the $p$-spin model with $p=4$,
the only difference being the presence of the Bessel functions instead of the hyperbolic
cosine in the integrals, the domain of integration in $z$ and a $z$ in
the integrand.

The equilibrium value of $m$ is the solution of 
$\partial_m \phi_{1RSB}=0$. 
At high temperature the solution $q=0$ and $m=1$ ({\it paramagnetic state}) is the stable one, 
while for $T<T_c$ a new solution with $q \neq 0$ and $m < 1$ ({\it spin glass}) becomes stable.
The temperature $T_c$ (also called Kauzmann temperature $T_K$)
marks the appearance of the thermodynamic glassy phase.

\subsection{Phase space structure of the model}

Starting from Eq.~(\ref{f1rsb}) one can repeat the analysis of \cite{Montanari03} to derive
the full phase space structure of the model at the {\sc 1rsb} level. Again, we will reproduce
in some details the original derivations for the reader who is not familiar with these methods.

In this class of mean field disordered models, at low temperature, the phase space in disconnected
in many {\it metastable states}, \ie local minima of the free energy.
The number of states of given free energy
density $f$ is $\O(f)=\exp N \Si(f)$. The function $\Si(f)$
vanishes continuously at $f=f_{min}$ and drops to zero above $f=f_{max}$ 
(see \eg \cite{CC05}).
The main peculiarity of these models is that an 
{\it exponential number} of metastable states is present at low 
enough temperature.

One can write the partition function $Z$, at low enough temperature and for $N \to \io$, 
in the following way:
\beq
\label{Zm1}
\begin{split}
Z = e^{-\b N F(T)} \sim \sum_\a e^{-\b N f_\a}
= \int_{f_{min}}^{f_{max}}df \, e^{N [\Si(f)-\b f]}
\sim e^{N [\Si(f^\star)-\b f^\star]} \ ,
\end{split}
\eeq
where $f^\star \in [f_{min},f_{max}]$ is such that $\Phi(f)=f - T \Si(f)$ 
is minimum, \ie it is the solution of
\beq
\frac{d\Si}{df} = \frac{1}{T} \ ,
\eeq
provided that it belongs to the interval $[f_{min},f_{max}]$.
Starting from high temperature, one encounters three temperature regions:
\begin{itemize}
\item For $T > T_d$, the free energy density of the paramagnetic state is
smaller than $f - T\Si(f)$ for any $f\in [f_{min},f_{max}]$, so the paramagnetic
state dominates (in this region the decomposition
(\ref{Zm1}) is meaningless).
\item For $T_d\geq T \geq T_c$, a value $f^\star \in [f_{min},f_{max}]$ is found, such that
$f^\star - T \Si(f^\star)$ is equal to $f_{para}$. This means that the paramagnetic state
is obtained from the superposition of an
{\it exponential number} of states of {\it higher} individual free energy
density $f^\star$. The phase space is disconnected in this exponential number of
regions: however, no phase transition happens at $T_d$ because of the
equality $f^\star - T \Si(f^\star)=f_{para}$ which guarantees that the free energy is
analytic on crossing~$T_d$. 
\item For $T < T_c$, the partition function is dominated by the lowest free 
energy states, $f^\star = f_{min}$, with $\Si(f_{min})=0$ and 
$F(T)=f_{min} - T \Si(f_{min}) = f_{min}$. At $T_c$ a phase transition occurs,
corresponding to the 1-step replica symmetry breaking transition found in the
replica computation.
\end{itemize}
In the range of temperatures $T_d > T > T_c$, the phase space of the model
is disconnected in an exponentially large number of states, giving a contribution
$\Si(T) \equiv \Si(f^\star(T))$ to the total entropy of the system.
This means that the entropy per particle $S(T)$ for $T_d > T > T_c$ can be written as
\beq
S(T) = \Si(T) + S_{vib}(T) \ ,
\eeq
$S_{vib}(T)$ being the individual entropy of a state of free energy $f^\star$.
The task is then to compute the function $\Si(f)$ at fixed $T$ and the
{\it equilibrium complexity} $\Si(T)=\Si(f^\star(T))$.

To this aim,
the idea \cite{Mo95} is to consider $m$ copies of the original system, 
coupled by a small attractive term added to the Hamiltonian.
The coupling is then switched off after the thermodynamic limit has been taken.
For $T<T_d$, the small attractive coupling is enough to constrain
the $m$ copies to be in the same state.
At low temperatures, the partition function of the replicated system is 
then
\beq
\label{Zm}
Z_m = e^{-\b N \Phi(m,T)} \sim \sum_\a e^{-\b N m f_\a}
= \int_{f_{min}}^{f_{max}}df \, 
e^{N [\Si(f)-\b m f]}
\sim  e^{N [\Si(f^\star)-\b m f^\star]} \ ,
\eeq
where now $f^\star(m,T)$ is such that $\Phi(m,f)=m f - T \Si(f)$ is minimum and
satisfies the equation
\beq
\frac{d\Si}{df} = \frac{m}{T} \ .
\eeq
If $m$ is allowed to assume real values by an analytical continuation, 
the complexity can be computed from the knowledge
of the function $\Phi(m,T)=m f^\star(m,T) - T \Si(f^\star(m,T))$. 
Indeed, it is easy to show that
\beq
\label{mcomplexity}
\begin{split}
&f^\star(m,T) = \frac{\partial \, \Phi(m,T)}{\partial m} \ , \\
&\Si(m,T) = \Si(f^\star(m,T)) = m^2 \frac{\partial \,[ m^{-1} \b \Phi(m,T)]}{\partial m} = 
m \b f^\star(m,T) - \b \Phi(m,T) \ .
\end{split}
\eeq
Thus the function $\Si(f)$ can be reconstructed from the parametric plot of $f^\star(m,T)$ 
and $\Si(m,T)$ by varying $m$ at fixed temperature.
The equilibrium complexity is simply $\Si(T)=\Si(m=1,T)$.

Using the replica trick to compute the free energy,
\beq
\Phi(m,T)=-\frac{T}{N}\overline{\log Z_m} = -\frac{T}{N}
\lim_{n\to 0} \frac{ \overline{ ( Z_m )^n }-1}{n} = -\frac{T}{N}
\lim_{n\to 0} \frac{ \overline{ Z^{mn} } -1}{n}  \ ,
\eeq
one obtains the partition function of $nm$ copies of the system, with the constraint
that each block of $m$ replicas has to be in the same state, \ie the replicas must have
nonzero overlap. This leads naturally
to the {\sc 1rsb} structure for the overlap matrix (with $m$ fixed), see Eq.~(\ref{QQ1rsb}), and
\beq\label{GGGG}
\Phi(m,T) =  -\frac{T}{N}
\lim_{n\to 0} [\exp \big[ -\b n m N \phi_{1RSB}(m,\overline{q},T) \big]-1]/n = 
m \phi_{1RSB}(m,T) \ .
\eeq
Note that the hypothesis that the $m$ replicas are in the same state implies that for
any value of $(m,T)$ one has to substitute in $\phi_{1RSB}$ the nonzero solution of Eq.~(\ref{q1rsb}),
$q^\star(m,T)$.
Above $T_d$ this solution disappears as a vanishing coupling cannot constrain the replicas to stay close
to each other.

Using Eq.s~(\ref{mcomplexity}) and (\ref{GGGG}) the complexity as a function of $m$ is
\beq
T \Si(m,T) = m^2 \partial_m \big[ m^{-1} \Phi(m,T) \big] = 
m^2 \partial_m \phi_{1RSB}(m,q^\star,T) \ ,
\eeq
and the equilibrium complexity is
\beq
\Si(T)=\Si(1,T)=-\frac{3 \b^2 (q^\star)^4}{4} + \ln \int_0^{\infty} {\cal D}z \ I_0(\b\l^\star z)
-\frac{\int_0^\io {\cal D}z \ I_0(\b\l^\star z) \ln I_0(\b\l^\star z)}{\int_0^{\infty} {\cal D}z \ I_0(\b\l^\star z) }
\ ,
\eeq
where $\l^\star = \sqrt{2 (q^\star)^3}$.

\subsection{Phase diagram in the $(m,T)$-plane}
\label{sec:risultati}

\begin{figure}
\centering
\includegraphics[width=.55\textwidth]{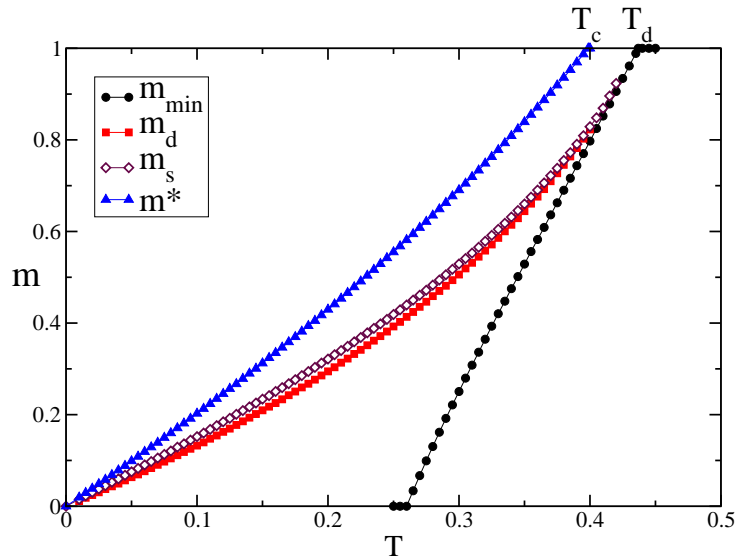}
\caption{(Color online) Phase diagram of the model in the $(m,T)$ plane (see Section IV D in the text).}
\label{emmi}
\end{figure}

For a given value of $T$, we can identify four relevant values of $m$. They are reported
in Fig.~\ref{emmi} and are defined as follows:
\begin{enumerate}
\item A solution $q^\star(m,T) \neq 0$ of Eq.~(\ref{q1rsb}) 
is present for $m \geq m_{min}(T)$. Thus, in the region $m \geq m_{min}(T)$,
$\Phi(m,T)$ is well defined and we can compute the complexity $\Si(m,T)$ and
the free energy $f^\star(m,T)$ using Eq.s~(\ref{mcomplexity}).
\item The value $m^\star(T)$ such that $\Si(m,T) =0$ correspond to the solution of
the thermodynamics and $f^\star(m^\star,T)=\phi_{1RSB}(m^\star,T)$ is the free energy 
in the spin glass phase. The temperature $T_c$ is defined by $m^\star(T_c)=1$.
\item The function $f^\star(m,T)$ has a maximum for $m=m_d(T)$. This means that $f^\star(m_d,T)$ is
the maximum possible free energy $f_{max}(T)$. The states with $f=f_{max}$ are called
{\it threshold states} \cite{CC05,Montanari03}.
\item Finally, one can investigate the stability of the {\sc 1rsb} solution with respect 
to further steps of replica symmetry breaking, following the analysis of \cite{Ga85}.
We report the details of the calculation in the Appendix \ref{App_C}.
It turns out that the {\sc 1rsb} solution is stable toward further steps of
replica symmetry breaking for $m \geq m_s(T)$.
\end{enumerate}

The thermodynamic phase diagram of the model is easily deduced from Fig.~\ref{emmi}.
The paramagnetic solution is stable for $T>T_c$. 
The temperature $T_c$
is the thermodynamic glass transition temperature, at which a
downward jump of the specific heat is observed, as in standard first-order transition.
Below this temperature the {\sc 1rsb}
spin glass solution is stable, and remains such down to $T=0$, as $m^\star(T) > m_s(T)$ for
all $T$. Thus in this model no Gardner transition (transition to a full {\sc rsb} solution \cite{Ga85})
is observed.

Some information on the dynamics of the model can also be obtained from Fig.~\ref{emmi}.
Indeed, at $T_d$ (defined by $m_d(T)=1$) a {\it dynamical transition} takes place \cite{CC05}:
correlation functions are expected to develop an infinitely long {\it plateau} 
and the system becomes dynamically trapped in one of the exponentially 
large number of states that appears at $T_d$, as discussed above.

\begin{figure}
\centering
\includegraphics[width=.55\textwidth]{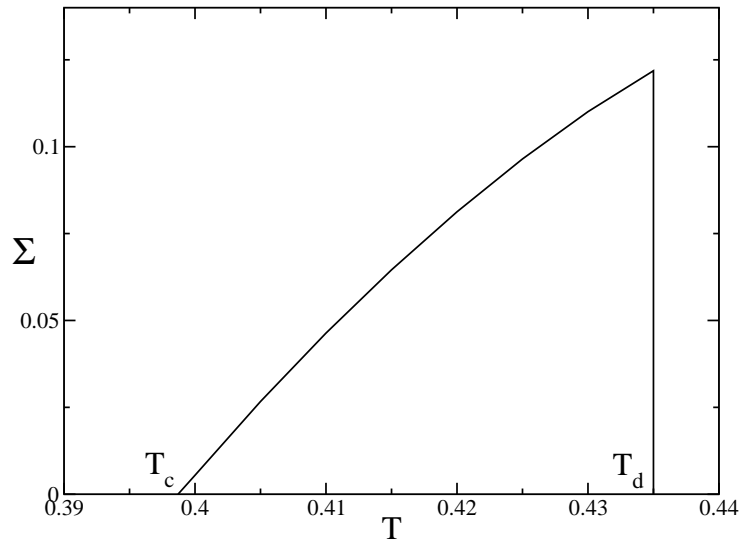}
\caption{The equilibrium complexity $\Si(T)$ as a function of the temperature.}
\label{fig:complexity}
\end{figure}

Between the static transition temperature $T_c$ and the dynamic one $T_d$ 
the phase space has a non-trivial shape: it is disconnected in an exponential
number of states ${\cal N}(T) = \exp N \Si(T)$,
where $\Si$ is the configuration entropy (or complexity) of the system.
In Fig.~\ref{fig:complexity} the quantity $\Si$ is reported as a function of $T$.
The point $T_c$ at which the complexity vanishes $\Si(T_c)=0$
signals the appearance of the thermodynamic phase transition.

However, it is not clear what is the dynamic behavior of the model if quenched from
$T=\io$ to $T < T_d$. Indeed, as already observed in the Ising $p$-spin glass model,
the threshold states always lie into the {\sc 1rsb}-unstable region, \ie $m_d < m_s$ for
$T<T_d$. This means that the {\sc 1rsb} ansatz is unable to give the correct prediction
for these states below $T_d$ and one need to consider further steps of
replica symmetry breaking. The investigation of the dynamics of the model after a quench
below $T_d$ will be the object of future investigation.

\section{Route to the experimental observation of the glassy transition}
The investigated model exhibits a dynamical transition at $T_d$ and 
a thermodynamic phase transition at a lower temperature $T_c$ 
(characterized by one-step replica symmetry breaking scenario),
very similar to $p$-spin glass models.\\
Let's turn our attention to the physical interpretation of these transitions.
We recall that the ``temperature'' $T$ introduced in the model is defined as the ratio between the 
{\it bath} temperature $T_{bath}$, measuring the optical noise in the system, and the pumping rate ${\cal P}$, 
so we can vary the latter to explore the phase diagram of the system.

Starting from low pump intensity (high temperature) and increasing ${\cal P}$
(decreasing $T$), different interesting phenomena take place.
The relevant quantities to look at in experiments are correlation functions in time domain,
i.e. the self correlation
functions of a specific frequency ($\omega_m$) component of the
electric field in the cavity (for example via heterodyne experiments, see below):
\begin{eqnarray}
C(t,\omega_m)= \langle a_m(t+\t) a^*_m(\t) \rangle_\t= A_m^2 \left \langle \exp \{ i [\varphi_m(t+\tau)-
\varphi_m(\tau)] \} \right
\rangle_{\tau} \ ,
\end{eqnarray}
where the $\langle \dots \rangle_{\tau}$ is the average over the time origin $\t$.
At high $T$, because of the fast dynamics of the phases, $C(t,\omega_m)$ decays to zero on short times.
On lowering $T$ the dynamics of phase variable $\varphi_m(t)$ becomes
slower and slower and $C(t,\omega_m)$ is expected to decay towards
zero in longer and longer times.
At the dynamic transition point $T_d$,
the dynamics of the $\varphi$'s becomes non-ergodic, they are no
longer able to explore the whole phase space and the function
$C(t,\omega_m)$ decays towards a plateau:
the mode's phases $\varphi_m(t)$ are locked
to some fixed random values (``random mode-locking'')
and oscillate around these values.
The fact that the complexity $\Si$ is different from zero at the dynamic transition
point implies that there is an exponentially large number of possible values for the
locked phases and,
correspondingly, many different time structures
of the electric field in the random laser.

More technically, the region between $T_d$ and $T_c$ is dynamically not-accessible,
due to the mean-field character of the interactions. 
Only for short range models, where activated processes become possible,
this interesting region can be explored.
We can expect that, if the mean-field approximation is not fully verified 
by the random lasers, see the discussion in section~\ref{sec:modes},
the system would be able to enter in the {\it activated} (or Vogel-Fulcher) regime (to use a terminology
familiar in the liquid-glassy physics) and the correlation functions decay to zero at long times 
after the plateau. In this region the relaxation time is found to scale as 
$\t = \t_0 \exp[D/(T-T_c)]$ and
only at the thermodynamic phase transition point $T_c$ the system is really locked in the {\it ideal}
glassy state (the relaxation time becomes infinite).
Similarly to what happens in $p$-spins and structural glasses,
interesting phenomena as aging, memory effects, and history dependent responses 
are expected to take place for $T_c < T < T_d$, see \eg \cite{YoungBook,Biroli05,Cu02} for
recent reviews.

It is worth to remark that in the region where these phenomena are expected to
happen, the relaxation time of the system is larger, by many orders of
magnitude, than the typical microscopic time scales. For instance, in molecular
glasses where the typical time scale is $\t_0 \sim 10^{-12}$s, the relaxation
time of the system can be as large as $100$s already for $T_g \sim 1.5
T_c$. These systems cannot be equilibrated close to $T_c$ because of the
exponential divergence of the relaxation time for $T\to T_c$.
In random lasers, the typical time scale for photon dynamics is $\t_0 \sim 10^{-14}$s,
so one will gain orders of magnitude in time. This should
allow to equilibrate the system closer to $T_c$. Even if the decrease in
$T-T_c$ that one can achieve in this way will not be substantial, it could be
enough to test the predictions of some recent theories of the glass transition
in presence of short range interactions \cite{YoungBook,Biroli05}.

Moreover, we recall that, as discussed in section~\ref{sec:modes}, in random
lasers it is possible, in principle, to tune the interaction range (by tuning
the localization length), and thus the relevance of the activated processes. 
This is similar to what has been done theoretically by considering the Kac 
limit for spin glasses \cite{Franz04}
and might allow to explore the crossover between the mean field and the activated regime and to shed light on
some debated issues concerning the nature of the spin glass phase in real systems.
Before concluding, it is important to provide an order of magnitude estimate of physical quantities involved 
in the experiments and to address some possible experimental frameworks.
\subsection{Order of magnitudes}
For the sake of concreteness we will focus, as an example, on 
recent experiments in random lasers realized
by scatterers (e.g. zinc oxide powder dispersed in a solvent doped with a dye, as e.g. in \cite{Mujumdar04}).
These experiments employed pumped source beams (see below), for the moment we show that
the glassy transition can be expected for the currently employed pump power levels.
We stress once again that our model is sufficiently general to embrace a much wider variety of disordered 
amplifying systems.

We start from the coupling coefficients, which are given by Eq.~(\ref{gs});
the fields are real valued numbers, their modulus scale as $|E|\cong 2V^{-1/2}/\sqrt{\epsilon_0 n_0^2}$ due to the normalization 
($n_0$ is an average refractive index);
their sign can be ``embedded'' in the sign of the $\chi$ coefficients; additionally $\omega_s\propto \omega_0\equiv2 \pi c/\lambda$.
Hence, omitting indexes, $\langle g\rangle=0$ and
\begin{equation}
g\simeq \frac{8\omega_0^2}{  V^2 \epsilon_0^2 n_0^4} \int_V \chi({\bf r}) dV
\end{equation}
We need $\langle g^2\rangle$, and we assume 
\begin{equation}
\langle \chi({\bf r}) \chi({\bf r}') \rangle=\chi_0^2 L_r^3 \hat \delta({\bf r}-{\bf r}')
\end{equation}
with $\chi_0$ a typical nonlinear susceptibility value, $L_r$ a characteristic length for the disorder 
and $\hat \delta$ the {\it coarse grained} Dirac delta. 
Thus
\begin{equation}
\langle g^2 \rangle \simeq \left ( \frac{\omega_0^2}{2 V^2 \epsilon_0^2 n_0^4} \right)^2 \langle \int_V \chi({\bf r}) dV \int_V \chi({\bf r'}) dV' \rangle=
\left ( \frac{8 \omega_0^2}{V^2 \epsilon_0^2 n_0^4} \right)^2 \chi_0^2 L_r^3 V
\end{equation}
We have then for the standard deviation
\begin{equation}
\label{stdg1}
\sqrt{\langle g^2 \rangle}\simeq \frac { 8 \omega_0^2 \chi_0 L_r^{3/2}} { \epsilon_0^2 n_0^4} \frac{1}{V^{3/2}}
\end{equation}
For the sake of simplicity, let us consider a box with volume $V$: the number of modes per unit of volume and unit 
of frequency is given by 
(this is just an estimate, as in nanostructured optical cavities the density of modes can be enhanced or depressed 
with respect to a standard box \cite{SakodaBook})
\begin{equation}
\rho(\nu)=\frac{8 \pi n_0^3 \nu^2}{c^3}
\end{equation}
hence for $N$ modes in a wavelength range $\Delta \lambda$ ($\lambda=c/\nu$) it is
\begin{equation}
N=\frac{8 \pi n_0^3 \Delta\lambda}{\lambda^4} V
\label{Nmodes}
\end{equation}
Eq. (\ref{Nmodes}) is used in (\ref{stdg1}) and gives
\begin{equation}
\sqrt{\langle g^2 \rangle}\simeq \frac {8^{5/2} \pi^{3/2}n_0^{1/2} \omega_0^2 \chi_0 L_r^{3/2}} 
{ \epsilon_0^2 n_0^4} \frac{(\Delta\lambda)^{3/2}}{\lambda^{6}}\frac{1}{N^{3/2}}\text{,}
\end{equation}
which scales as $N^{-3/2}$ as anticipated. Next we have to consider the coefficients $G_{spqr}=g_{spqr} A_s A_p A_q A_r$:
\begin{equation}
\sqrt{\langle G^2 \rangle}\simeq 
\frac {8^{5/2} \pi^{3/2} \omega_0^2 \chi_0 L_r^{3/2} n_0^{1/2}} { \epsilon_0^2 } \frac{(\Delta\lambda)^{3/2}}{\lambda^{6}}\frac{\langle A^2 \rangle^2}{N^{3/2}}\text{,}
\end{equation}
and the coefficients $J=G/(g_0 \langle A^2 \rangle^2)$ with variances $8/N^3$. We can hence determine $g_0$ as
\begin{equation}
\label{g0}
g_0 \cong \frac {8 \pi^{3/2} n_0^{1/2} \omega_0^2 \chi_0 L_r^{3/2}} { \epsilon_0^2 } \frac{( \Delta\lambda)^{3/2}}{\lambda^{6}}
\end{equation}
and, finally, the adimensional $\beta=1/T$ is given by
\begin{equation}
\label{beta}
\beta \cong \frac {8 \pi^{3/2} \omega_0^2 \chi_0 L_r^{3/2}} { \epsilon_0^2 } 
\frac{( \Delta\lambda)^{3/2}}{\lambda^{6}}\frac{\langle A^2 \rangle^2}{k_B
  T_{bath}} \ .
\end{equation}
Remembering that the average energy per mode is $\omega_0 \langle A^2\rangle$, we obtain, at the transition,
the threshold value 
(denoting $T_d \sim 0.435$ the normalized temperature at the dynamic
transition, see Fig.~\ref{fig:complexity}):
\begin{equation}
\label{ersbenergy}
\mathcal{ E}_{RSB}=\omega_0 \langle A^2\rangle=\sqrt{\frac{k_B T_{bath}
    \epsilon_0^2 \lambda^6}
{8 \pi^{3/2}n_0^{1/2} T_d\chi_0 (L_r \Delta \lambda)^{3/2}}}\text{.}
\end{equation}
The noise temperature can be taken as due to the spontaneous emission, which is typically the dominant contribution: 
following \cite{YarivBook}, $2k_B T_{bath}=\hbar (N_2/(N_2-N_1))_t/\tau\cong \hbar/\tau$, with $\tau$ the average life time per mode, 
and $(N_2/(N_2-N_1))_t$ the population inversion at lasing threshold.
Taking typical values from the reported experiments, 
($\Delta \lambda=100$ nm, $\lambda=630$ nm, $n_0=2$,$L_r=10$ nm, $\tau=100$ fs) and for 
the susceptibility $\chi_0=10^{-27}$ CmV$^{-3}$ \cite{BoydBook} it is $\mathcal{ E}_{RSB}\cong10^{-16}$ J. 
Assuming a pumping beam with peak power $ P_{RSB}\cong N\mathcal{ E}_{RSB}/\tau$,  
gives $ P_{RSB}\cong 0.1$ W with $N=100$, which focused on the typical area of $100$ $\mu$m$^2$ provides the typical 
values for the peak pump intensities used in the experiments ($\simeq 100$ kW/cm$^2$). 
Thus we expect that the ``glass transition'' can be observed within the currently available experimental
framework.

\subsection{Continuous-wave random lasers}
All the theory reported in this manuscript makes reference to continuous-wave (CW) RLs, 
hence we start discussing this kind of systems. 
Experimental investigations of CW-RL were already reported in nano-powders (see \cite{Redmond04} and references therein);
alternative experimental geometries include disordered photonic crystals \cite{SakodaBook}, as membranes or multi-layered
systems with gain provided e.g. by quantum wells in semiconductor materials.
In these integrated high-index contrast geometries multi-mode random-cavity lasers can in principle operate in CW regime, 
and this opens the way to a comprehensive experimental analysis of the dynamics of the laser emission. 
The experimental setup can follow previous investigations of the 
noise figure of standard semiconductor lasers (see e.g. \cite{Vahala83}).
RL emission is collected, and filtered in a narrow band 
in order to select one or few modes.  
At the mode locking transition the intensity signal $I(t)$ is expected to switch from a 
random noise superimposed to a CW value, to a largely modulated line-shape displaying a random 
sequence of disordered pulses. 
Heterodyne measurements are employed to extract phase and amplitude noise 
from which the dynamics of the amplitudes of the modes and their phase are extracted, as
well as the coherence function $C(t,\omega)$.
As discussed above, the fact that amplitude fluctuations make a negligible
contribution to the field autocorrelation is well known from laser theory \cite{Vahala83}, hence
$C(t,\omega)$ gives information on the phase-dynamics.

\begin{figure}
\includegraphics[width=15cm]{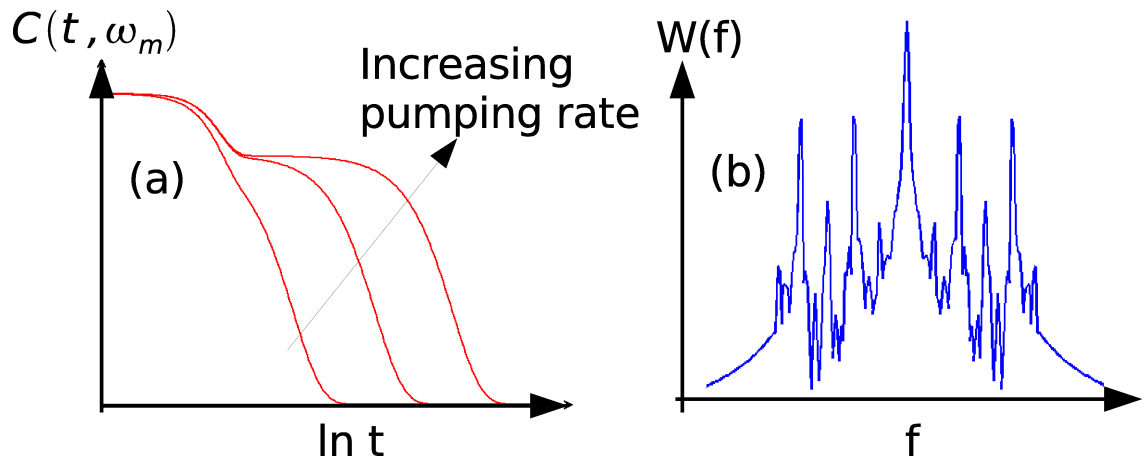}
\caption{
Sketch of two experimental signatures of the onset of a glassy transition of light in random lasers: (a) The mode
coherence function develops a plateu at high pumping rates, denoting ergodicity breaking. (b) The power density spectrum 
of the mode instantaneous frequency displays modulation sidebands; this corresponds to 
the transition from random-mode-phases to phase-locking in one the  meta-stable states; the side-bands are due to 
small oscillations in these minima that result into frequency shifts of the mode resonances.
\label{figsetup}}
\end{figure}

Specifically, before the glassy transition (low pumping rate) the line-width of the laser modes is wide and the autocorrelation $C(t,\omega)$ of
the mode signal is expected to decay with a single exponential time constant, corresponding to a Lorentzian line-shape
of the noise spectrum of the field. 
At the glass transition the laser dynamics slows down, and this result into a slower decaying of $C(t,\omega)$,
with the appearance of multiple time scales, and eventually to 
an ergodicity breaking corresponding to a plateau in the $C(t,\omega)$ signal (as sketched figure \ref{figsetup}a).

Further information is retrieved by phase demodulation  (e.g. by employing the usual combination of a limiter and a discriminator \cite{Vahala83})
whose output is the instantaneous frequency $d\phi_m /dt$, whose power density spectrum can be determined
by a spectrum analyzer. When the phase of the filtered modes are locked, they oscillate around one of the many equilibrium values.
Correspondingly, the spectrum of the phase noise for each mode is expected to switch from a wide line
to a narrow one displaying modulation side bands, which are due to the fact that the phases display
small oscillations around one of the many phase-locked states (as sketched figure \ref{figsetup}b).
\subsection{Pulsed random lasers and speckle patterns}
Most of the reported experiments on RLs have been done by using pumped laser beams,
with pulse duration from tens of picoseconds to tens of nanoseconds. 
One could argue if the mentioned mode-locking transition can be actually observed in these regimes.

First of all we observe that our theory deals with the phase-dynamics of the (quasi-)modes of a RL.
As far as the laser reaches a steady state for the amplitudes (and this is expected to happen in the leading
edge of a nano-second pump pulse, taking into account typical lifetimes; see e.g. \cite{Deych05} for
an extensive discussion of the mode amplitude  dynamics), the phases are expected 
to vary on the $\t_0 \sim 10$ fs time-scale. Indeed, in the framework of the Lamb's two level theory, they are affected by the dynamics
of the resonant medium polarization, i.e. by the time-constant of the off-diagonal density matrix elements,
whose inverse is the  ``dipole dephasing rate'' which is around $10^{14} \ \text{s}^{-1}$ (see e.g. \cite{Sierks98}). 
The latter time scale is the ``elementary'' time scale for the dynamics of the
phases, that corresponds in molecular systems to the typical time scale of
atomic vibrations $\t_0 \sim 10^{-12}$ s.

In a first approximation, the arrival of a pulse produces a fast variation of
the ``temperature'' from $T \sim \io$ (before the pulse) to a value of 
$T<T_d$ (subsequently after the arrival of the pulse). This corresponds, in the
language of molecular glasses, to an instantaneous (i.e. on the scale of
$\t_0$) {\it quench} of the system from infinite (or very high) temperature to
below $T_d$. On a very general ground, it is expected that the system will then
start to {\it age} \cite{Cu02,YoungBook}, in the sense that the
relaxation time $\t$ of $C(t)$ will increase with the time $t_w$ elapsed after the
arrival of the pulse (the quench). For systems that are in the class of the
$p$-spin model it is generally found that, for $t_w \gg \t_0$, $\t(t_w) \sim t_w$ (see as a
striking example Fig.~5 in \cite{KB97}). This means that for a pulse duration
$t_w \gg \t_0 \sim 10$ fs, the relaxation time
of the phases will be of the same order of the pulse duration and they
will appear to be frozen on this time scale.
Hence the mode-locking process should be observable in standard random laser for $ns$ pump pulses.

Note that, incidentally, the validity of the CW approximation has been
thoroughly discussed in a recent paper for nano-second pulsed laser \cite{derMolen06}.

Additionally, the pulsed regime favors the investigation of the correlation between different laser shots with fixed
disorder. In the presence of many modes, due to mode interferences and complex phase modulations, the peaks in the spectrum are
expected to largely vary from pulse to pulse.
This is also due to the fact that with a pulsed pump, in a thermodynamic language, the system is first  ``cooled''
(i.e. the average energy per mode is increased as the pump power increase) and then  ``heated'' (as in the trailing
edge of the pump pulse) \footnote{We thank an anonymous referee for addressing
  this point}.
Between two subsequent pulses the system is at infinite temperature and will
rapidly loose memory of the previous state:
as a result, in the presence of an exponentially large number of thermodynamically equivalent states, the system settles in a different
minima from pulse to pulse and correspondingly the phases will largely vary from shot to shot.
During the laser oscillation, the phase-modulation corresponding to the mode-locking process should be visible.
Similar phenomena (namely a large variation of the emitted spectrum from pulse to pulse) 
were already reported in the literature (as in Ref. \cite{Mujumdar04}).

A note on the speckle pattern of the emitted light is in order.
Indeed the speckle pattern is determined by the phase difference between the modes,
hence the spatial distribution of the emitted light is expected to largely vary from one shot to another of the random laser, when
the pump power is above the threshold for the glass transition.
It is important to emphasize that other authors predicted an exponentially large number of speckle 
patterns in nonlinear random media \cite{Spivak04}, however, in that case, the leading mechanism was the 
non-resonant Kerr effect (incidentally, our model Eqs.(\ref{H4}) also applies for
Kerr media as will be discussed elsewhere).
\section{Conclusions}
We derived a statistical model for the mode phases in a random laser. 
The obtained Hamiltonian resembles that of some spin glass model (the $p$-spin model with $p=4$
\cite{Montanari03,Ga85,Crisanti05,Nieuwenhuizen95}), 
and can also be thought as a generalization to the disordered case of a toy model 
Hamiltonian recently studied (the $k$-trigonometric model \cite{Angelani2003}).
The relevant parameter for exploring the phase diagram is a scaled
temperature $T$, the ratio between the ``true'' bath temperature and the square pumping rate
(the energy stored on average in each mode). 
Using standard statistical mechanics techniques of disordered systems
(i.e. replica method), we predict the existence of a dynamic transition 
at $T_d$ and of a thermodynamic phase transition at $T_c$,
characterized by one-step symmetry breaking scenario.
Between the two  temperatures, the appearance of an exponentially large number of states
is expected.
This corresponds to the existence of a
{\it random-mode locking} transition in random lasers:
looking at self-correlation functions of a specific frequency component of the electric field 
in the cavity, one should observe a non-ergodic behavior at $T_d$, 
i.e. the decay towards a {\it plateau}, where  
the phases are locked at random values.
The mode-locking will happen in a configuration of the phases
that depends on the history for a given sample, because the system will reach
a different metastable state depending on the initial state, with large 
sample-to-sample fluctuations. Thi can be observed by looking at the frequency fluctuation spectrum or
at the speckle pattern of the emitted light.

The logarithm of the number of these possible random configurations of the
phases is given by the complexity (see Fig.~\ref{fig:complexity})
times the number of active modes.
As long as the physical realization of the random laser is well described by the mean-field 
Hamiltonian, the system is not dynamically able to explore the phase space for 
temperature $T<T_d$ (or pumping rate higher than that corresponding to the dynamic transition)
and remains trapped always in the state reached at $T_d$.
However, taking into account the fact that the mean-field scenario could be a too ``crude''
approximation for real random lasers, 
we expect that the system will be able to explore on a long time scale
 $T_c<T<T_d$ region, where aging, memory effects, and
history dependent responses are expected to take place.

The expert reader in nonlinear optics or Bose-Einstein condensation will certainly recognize in our model a typical
system for many-modes interaction processes (e.g. solitons, parametric processes, supercontinuum generation ...); we believe indeed
that our result are also relevant in many branches of modern nonlinear physics involving disordered systems. 
Spin glasses have been defined has the ``most complex kind of condensed state'' \cite{FHbook},
we are convinced that there is no difficulty in accepting the emission of random lasers
as the ``most complex kind of light''.

\acknowledgments

It is our pleasure to thank many colleagues who expressed their interest in this work and helped us to improve the 
presentation of our work, by not refraining from many discussions, suggestions
and detailed criticisms: T.~Castellani, S.~Cavalieri, C.~Flytzanis, D.~Huse, J.~Kurchan, L.~Leuzzi,
R.~Livi, G.~Oppo, A.~Politi, F.~Ricci-Tersenghi, D.~Wiersma.
F.Z. is supported by the EU Research Training Network STIPCO (HPRN-CT-2002-00319).

\appendix

\section{Equations for the amplitudes and for the phases}
\label{App_0}

Here we will show that the Hamiltonian (\ref{Ham}) can be derived also directly
from the equation of motion for the complex amplitudes $a_s$, Eq.~(\ref{coupledmodes0}),
by assuming that the dynamics
of the phases is much faster than that of the $A_s$.

After (\ref{coupledmodes0}), using $g_{spqr}=g_{spqr}^{R}+i g_{spqr}^{I}$ and
$\eta_s=\eta_s^{R}+i\eta_s^{I}$, the equations for the
amplitudes are:
\begin{equation}
\label{amplitudes}
\begin{array}{l}
\displaystyle\frac{dA_s}{dt}=-\frac{1}{2}\sum_{pqr}A_p A_q A_r  \left[g_{spqr}^{R} 
\cos(\varphi_s+\varphi_p-\varphi_q-\varphi_r)+
g_{spqr}^{I} \sin(\varphi_s+\varphi_p-\varphi_q-\varphi_r)\right]+\\
+(\gamma_s-\alpha_s)A_s+\eta_s^{R}\cos(\varphi_s)+\eta_s^{I}\sin(\varphi_s)
\end{array}
\end{equation}
By assuming the phases as rapidly varying with respect to the amplitudes, these can be averaged out and the
``free run approximation''
 equations \cite{Lamb64,Bryan73,Deych05} are retrieved, providing the average energy in each mode $\mathcal{E}_s$:
\begin{equation}
\frac{d\mathcal{E}_s}{dt}=2(\gamma_s-\alpha_s)\mathcal{E}_s-\frac{ g_{ssss}^{R}}{\omega_s}\mathcal{E}_s^2+
\mathcal{E}_s \sum_r \frac{g_{srsr}^{R}+g_{ssrr}^{R}}{\omega_r} \mathcal{E}_r \ .
\end{equation}
Note that Eqs. (\ref{amplitudes}) take into account random cross- and self-saturation effects, by the terms
weighted by $g_{srsr}^R+g_{ssrr}^R$ and $g_{ssss}^R$ respectively, as well as the fact that the 
decay rates and gains are expected to be different for each mode. 
We model this circumstances in the text by taking the amplitude dependent $G$-coefficients as gaussianly distributed.

Next, we find the ruling equation for the phases after (\ref{coupledmodes0}):
\begin{equation}
\label{phase0}
A_s\frac{d\varphi_s}{dt}=-\frac{1}{2}\sum_{pqr} A_p A_q A_r 
\{g_{spqr}^{I} \cos(\varphi_s+\varphi_p-\varphi_q-\varphi_r)
-g_{spqr}^{R} \sin(\varphi_s+\varphi_p-\varphi_q-\varphi_r)\}
+\eta_s^{I}\cos(\varphi_s)-\eta_s^{R}\sin(\varphi_s) \ .
\end{equation}
Denoting  $\mathcal{E}_0=\omega_0 \langle A^2 \rangle$ the average energy per mode, we multiply (\ref{phase1}) by $\omega_0 A_s$ and take
$\omega_0 A_s^2\cong \mathcal{E}_0$ (while being $\omega_s \cong \omega_0$, as outlined above). 

The equation for the phases (\ref{phase0}) 
are equivalent to the following
\begin{equation}
\label{phase1}
\frac{d\varphi_s}{dt}=-\frac{\partial H_\varphi}{\partial \varphi_s}+\eta_s^{(\varphi)}
\end{equation}
with
$\eta_s^{(\varphi)}\equiv[ \eta_s^{I}\cos(\varphi_s)- \eta_s^{R}\sin(\varphi_s)]/A_s$ and
\begin{equation}
\label{Hcomplex0}
H_\varphi=\frac{1}{8}\sum_{spqr} \frac{\omega_0 A_s A_p A_q A_r}{\mathcal{E}_0} 
[g_{spqr}^{I}\sin(\varphi_s+\varphi_p-\varphi_q-\varphi_r)
+g_{spqr}^{R}\cos(\varphi_s+\varphi_p-\varphi_q-\varphi_r)]\ ,
\end{equation}
where we exploited the symmetries for the real part which are the same as discussed above, 
and where the imaginary part satisfy $g^{I}_{spqr}=-g^{I}_{qrsp}$.

Eq.~(\ref{Hcomplex0}) is cast in the form $H_\varphi=\omega_0 H/2\mathcal{E}_0$ with
\begin{equation}
\label{Hcomplex}
H=\frac14 \sum_{spqr} A_s A_p A_q A_r [g^R_{spqr} \cos(\f_s+\f_p-\f_q-\f_r) -
g^I_{spqr} \sin(\f_s+\f_p-\f_q-\f_r)] =
\frac{1}{4}\re\left[\sum_{spqr} g_{spqr} a_q a_r a_p^* a_s^*\right] \ .
\end{equation}
Eq.~(\ref{phase1}) is a Langevin system for the phases, and being (due to the fact that the noise $\eta$ is assumed to vary
on a much faster scale than the phases $\varphi$) 
$\langle \eta_p^{(\varphi)}(t) \eta_q^{(\varphi)}(t') \rangle= \omega_0 k_B T_{bath}
\delta_{pq}\delta(t-t')/\mathcal{E}_0$, 
its invariant measure is $\exp(-2 H_\varphi \mathcal{E}_0 /\omega_0 k_B T_{bath})=\exp(-H /  k_B T_{bath})$,
which is identical to the previous one. In the case of a complex coupling the Hamiltonian
will include also the second term in (\ref{Hcomplex}).
The replica analysis of the generalized model is a 
generalization of the one concerning Eq.~(\ref{hfinal0}) 
(see section \ref{replica_section}) and provides the same outcome for what concerns the existence of a RSB transition. 

\section{Self-consistency equation of 1RSB solution}
\label{App_B}
Here we derive the self-consistency equation for $q$, Eq.~(\ref{q1rsb}).
It is obtained imposing that the derivative of the free energy in Eq.~(\ref{f1rsb}) with respect to $q$ vanishes,
$\partial_q \b\phi_{1RSB}(m,q)=0$:
\beq
3\b^2q^2 [(1-m)q-1] =
- \frac{1}{m}\frac{\int_0^{\infty} {\cal D}z \ \partial_q I_0^m(\b\l z) }
{\int_0^{\infty} {\cal D}z \ I_0^m(\b \l z)} \ .
\label{eq1_a}
\eeq
Now, $\partial_q I_0^m = m I_0^{m-1} \partial_q I_0$ and 
$\partial_q I_0 =(\partial_\a I_0) (\partial_q \a)=I_1  \partial_q \a$, 
where $\a=\b\l z = \b \sqrt{2} q^{3/2} z$.
Then we can write:
\beq
m^{-1} \int_0^{\infty} {\cal D}z \ \partial_q I_0^m = 
\frac{3\b q^{1/2}}{\sqrt{2}}
\int_0^{\infty} {\cal D}z \ z I_0^{m-1} I_1 = 
\frac{3\b q^{1/2}}{\sqrt{2}}
\int_0^{\infty} dz e^{-z^2/2}\ \partial_z (z I_0^{m-1} I_1)  \ ,
\eeq
having used the identity $z e^{-z^2/2}=-\partial_z e^{-z^2/2}$ and integrating by part.
Performing the derivatives and using $\partial_z  =(\partial_z \a) \partial_\a  $
and the property of Bessel functions $\partial_\a I_1=I_0-I_1/\a$, we have:
\beq
m^{-1} \int_0^{\infty} {\cal D}z \ \partial_q I_0^m = 
3\b^2 q^2
\int_0^{\infty} {\cal D}z \ I_0^{m} \left[ 1+(m-1)\frac{I_1^2}{I_0^2}\right] \ .
\eeq
Substituting in Eq.~(\ref{eq1_a}) we then obtain the self-consistency equation:
\beq
q=\frac{\int_0^{\infty} {\cal D}z I_0^m(\b\l z) \left[\frac{I_1(\b\l z)}{I_0(\b\l z)}\right]^2}
{\int_0^{\infty} {\cal D}z I_0^m(\b \l z)} \ .
\label{B4}
\eeq

\section{Stability of 1RSB solution}
\label{App_C}

In this Appendix we discuss the stability of the {\sc 1rsb} solution.
First we have to compute
the Hessian of $\phi(q)$ evaluated in a solution that verifies the saddle point
equations $q_{ab}= \langle \cos(\f^a-\f^b)\rangle$, where we assume that $q_{ab}$ is
real. This assumption is motivated by the analysis of \cite{BiYo86}, p.894,
where it is shown for the case of a two spin interaction that even the full RSB
solution verifies $q_{ab}$ real for all $ab$. Still this is an assumption as
in principle there could be solutions such that $q_{ab}$ has an imaginary part
for $a\neq b$, see again \cite{BiYo86}.

Considering a perturbation $\d q_{ab}$ 
around the solution, one has, differentiating
Eq.~(\ref{f2min}):
\beq
G_{ab,cd} = \frac{d^2 \b\phi}{dq_{ab}dq_{cd}} = \frac{6\b^2}{n}  q_{ab}^{2}
\left[ \d_{ab,cd} - 
6\b^2 q_{cd}^{2} 
\big[\langle \cos(\f^a-\f^b)\cos(\f^c-\f^d) \rangle - q_{ab}q_{cd} \big] \right] \ .
\eeq
The condition $q_{ab}=q_{ba}^*$ implies $\d q_{ab} =\d q_{ba}$. 
The matrix $G$ is symmetric under the exchanges $a\leftrightarrow b$, $c\leftrightarrow d$.
When the matrices $G$ 
is evaluated in the {\sc 1rsb} solution, 
it is easy to see
that it becomes a block matrix that has non-vanishing elements only if
$(ab)$ and $(cd)$ belong to the same diagonal block related to one of the
diagonal blocks of the matrix $q_{ab}$. Thus we can restrict to consider
perturbations of one single block. With this restriction, substituting the
{\sc 1rsb} structure of $q_{ab}$, and neglecting the irrelevant prefactor
$6\b^2n^{-1} q^{2}$, $G$ has the following elements:
\beq
\begin{split}
&P\equiv G_{ab,ab} =  1
 -3\b^2 q^{2} \left[ 1 +
\frac{\int {\cal D}\z I_0(\b\l|\z|)^{m-2} I_2(\b\l|\z|)^2}
{\int {\cal D}\z I_0(\b \l |\z|)^m} - 2 q^2 \right] \ , \\
&Q\equiv G_{ab,ad} =
 -3\b^2 q^{2} \left[q+
\frac{\int {\cal D}\z I_0(\b\l|\z|)^{m-3} I_2(\b\l|\z|) I_1(\b\l|\z|)^2 }
{\int {\cal D}\z I_0(\b \l |\z|)^m} - 2 q^2 \right] \ , \\
&R\equiv G_{ab,cd} =
 -6\b^2 q^{2} \left[
\frac{\int {\cal D}\z I_0(\b\l|\z|)^{m-4} I_1(\b\l|\z|)^4 }
{\int {\cal D}\z I_0(\b \l |\z|)^m} - q^2 \right] \ .
\end{split}
\eeq
Following the analysis of \cite{deAlmeida78}, the relevant eigenvalue of $G$ that eventually
becomes unstable is $\L = P - 2Q +R$, \ie the stability condition is
\beq
\L = 1 -6\b^2 q^{2}
\frac{\int {\cal D}\z I_0(\b\l|\z|)^m \left\{
\frac{1}{2}\left[ 1
- \frac{I_1(\b\l|\z|)^2}{I_0(\b\l|\z|)^2} \right]^2 +
\frac{1}{2}
\left[ \frac{I_2(\b\l|\z|)}{I_0(\b\l|\z|)}
- \frac{I_1(\b\l|\z|)^2}{I_0(\b\l|\z|)^2} \right]^2 \right\}}
{\int {\cal D}\z I_0(\b \l |\z|)^m} > 0 \ .
\eeq
that is
\beq
\frac{1}{6\b^2  q^{2}} > 
\frac{\int {\cal D}\z I_0(\b\l|\z|)^m \left\{
\frac{1}{2}\left[ 1
- \frac{I_1(\b\l|\z|)^2}{I_0(\b\l|\z|)^2} \right]^2 +
\frac{1}{2}
\left[ \frac{I_2(\b\l|\z|)}{I_0(\b\l|\z|)}
- \frac{I_1(\b\l|\z|)^2}{I_0(\b\l|\z|)^2} \right]^2 \right\}}
{\int {\cal D}\z I_0(\b \l |\z|)^m} \ .
\eeq
Once the solution of the saddle point Eq. (\ref{B4}) is substituded in the expression above, 
one obtains the condition $m>m_s(T)$, see Fig. \ref{emmi}.

\end{document}